\documentclass[a4paper]{article}
\usepackage[ansinew]{inputenc}
\usepackage{times}
\usepackage[T1]{fontenc}
\usepackage{graphicx}
\usepackage{geometry}
\usepackage{amssymb}
\usepackage{amsmath}

\usepackage{rotating}

\usepackage{xcolor}
\colorlet{shadecolor}{gray!25}
\usepackage{subfig}
\usepackage{float}

\author{Ludwig A. Hothorn,\\ 
Im Grund 12, D-31867 Lauenau, Germany\\ \scriptsize(retired from Leibniz University Hannover)\normalsize\\
Dimitrios Spiliotopoulos, \\Xenometrix AG, CH-4123 Allschwil, Switzerland}

\title{Similarity of multiple dose-response curves in interlaboratory studies in regulatory toxicology}
\begin{document}

\maketitle
\begin{abstract}
To claim similarity of multiple dose-response curves in interlaboratory studies in regulatory toxicology is a relevant issue during the assay validation process. Here we demonstrated the use of dose-by-laboratory interaction contrasts, particularly Williams-type by total mean contrasts.
With the CRAN packages \textit{statint} and \textit{multcomp} in the open-source software R the estimation of adjusted p-values or compatible simultaneous confidence intervals is relatively easy. The interpretation in terms of global or partial equivalence, i.e. similarity, is challenging, because thresholds are not available a-priori. This approach is demonstrated by selected in-vitro Ames MPF assay data.
\end{abstract}

\section{Introduction}\label{sec1}
A relevant objective of interlaboratory studies in regulatory toxicology is to demonstrate the similarity of multiple dose-response curves in $l$ participating laboratories $Lab_l$. Almost all of these bioassays are based on the design $[C-, D_1, ...,D_k]$, assuming a primary normally distributed endpoint. This provides a comparison of dose-response curves, whereby the dose can be modeled as a qualitative factor using contrast tests or as a quantitative covariate using nonlinear regression models. Both approaches have advantages and disadvantages in this specific design (comparisons to C- are meaningful, $k$ is low, e.g. 3), see for details e.g. \cite{Hothorn2016}. Similarity is statistically translated into equivalence, which is demonstrated by inclusion in $(1-2\alpha)$ confidence intervals \cite{Bauer1996}. The crux of the matter is that the required tolerability limits are a-priori endpoint/assay specific not defined. Either a post-hoc comparative interpretation of the estimated intervals with regard to their tolerability is carried out or, alternatively, the proof-of-hazard standard used in toxicology is used as an equivalent, where $ p>0.10$.
One can demonstrate equivalence for a nonlinear function and thus the whole curve, or only a part of the curve (e.g. the linearized one) or for a single estimator, such as benchmark dose (BMD), LD50 or no-observed-adverse-effect-concentration (NOAEC).\\
Here we are guided by the evidence of a significant trend as an effectiveness criterion in the guidelines, e.g. for in-vivo micronucleus assay \cite{OECD474}. And as a trend test we choose the Williams test (\cite{Williams1971}) because it is geared to comparisons with C-, as a multiple contrast test it is robust for several forms of dose-response dependence, modeling the dose as a qualitative factor level and recommended in guidance, e.g. US-NTP for continuous endpoints. If the dose factor is modeled together with the laboratory factor, a lack of two-way interaction is a criterion for similarity. However, this global criterion cannot indicate that at least one, any interaction exists, but it cannot indicate which one exactly. It could be that 1 out of 10 laboratories behave differently (but the other 9 are similar). Or only the difference between $D_3-D_2$ is different in at least a single lab, which is toxicologically not relevant. Furthermore, confidence intervals are not readily available for the F-test statistics. A better alternative are interaction contrasts. For the factor dose the Williams contrast is described here, for the factor laboratory the total mean contrast (to leave the number of comparisons at $k$); other definitions are possible.\\

To characterize the Ames fluctuation test \cite{Reifferscheid2012} used the lowest ineffective concentration as criterion, the fold change increase to compare standard miniaturized and Ames II and MPF Assay \cite{Flick2012, Spil2020}, where the confidence intervals for benchmark dose estimates were used for the blood Pig-a gene mutation assay \cite{Dertinger2020} (requiring the same underlying nonlinear model for any condition).

\section{A motivating example: Ames MPF in-vitro assay}
\begin{figure}
	\centering
		\includegraphics[width=0.56\textwidth]{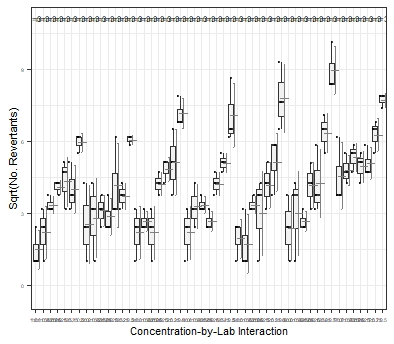}
	\caption{Boxplot for concentration-by-laboratory interaction in compound 4}
	\label{fig:IA4}
\end{figure}

In an extension interlaboratory study, the new MPF-assay was considered for 6 selected compounds, in 7 particular laboratories, using 5 strains and 2 kinds of metabolization \cite{Spiliotopoulos2020}. Substance 4 was used as data example, where $n_i=6$ (pooled over $n_i=3$ per metabolization S9+, S9-) per concentration and laboratory were available. The number of revertants were transformed to Nishiyama \cite{Nishiyama2003} to achieve approximate normal distribution in this small sample design. The boxplots in Figure \ref{fig:IA4} reveal monotonic increasing curves with similar shapes in the different labs, still additive shifts between the labs (e.g. all comparable values in lab 3 are less than in lab 7), tied values, variance heterogeneity (commonly higher variance in higher concentrations, but not always), and small sample sizes per concentration and laboratory (pooled over $n_i=3$ for S9+, S9-). 

\section{Williams-by-total mean interaction contrasts}
In the following, the interaction contrasts are derived for the two-way layout dose-by-laboratory. First, the Williams-contrast test is derived for the primary factor 'dose', then the contrast test for comparison with the total mean for the secondary factor 'laboratories', and finally the resulting interaction contrast.
 
\subsection{Factor dose: The Williams-type multiple contrast test}
Williams original procedure \cite{Williams1971} is a rather complex approach based on maximum likelihood estimators under total order restriction, hard to generalize. Here we use its re-formulation as multiple contrast test \cite{Bretz2006}, assuming order restriction with respect to C-. A multiple contrast test is a maximum test on $q$ elementary tests (where $q$ depends on the kind of multiple contrast) (i.e. a special version of an union-intersection test): $t_{MCT}=max(t_1,...,t_q)$ which follows jointly  $(t_1,\ldots,t_q)^\prime$ a $q$-variate $t$- distribution with a common degree of freedom $df$ and the correlation matrix  $R$, which is depending on $c_i, n_i$ under simple assumptions (or more complex for other MCT's \cite{Hothorn2016}).  The included single contrast tests $t_{SC}=\sum_{i=0}^k c_i\bar{x}_i/S\sqrt{\sum_i^k c_i^2/n_i}$ and $\sum_{i=0}^k c_i=0$ differ by their weights $c_i$, i.e. particular contrast matrix definitions determine the particular test version. For the simple balanced design with k=3 the contrast matrix for the one-sided Williams-type test is:

   \begin{table}[htbp]
            \begin{tabular}{ r c c c c }
         $c_i$ & C & $D_1$ & $D_2$ &$D_3$ \\
        $c_a$ & -1 & 0   & 0 & 1   \\
        $c_b$ & -1 & 0 & 1/2 & 1/2  \\
				$c_c$ & -1 & 1/3 & 1/3 &1/3  \\
				\end{tabular}
        \end{table}
The simple Williams contrast illustrates the idea and interoperability of an MCT: either the comparison of $D_3$ vs. C is significant (i.e. strict monotone), or to the pooled $(D_3+D_2)/2$, or $(D_3+D_2+D_1)/3$, i.e. a plateau (or all three, or neither ($H_0$).  Either multiplicity-adjusted p-values or simultaneous confidence limits (here one-sided lower limits) are available:
    $[\sum_{i=0}^k c_i\bar{x}_i - S* t_{q,df,R,2-sided,1-\alpha}\sqrt{\sum_i^k c_i^2/n_i}]$.\\
		
This makes the Williams test the recommended trend test in pharmacology/toxicology: sensitive to some monotonic and partially non-monotonic forms, a comparison to the control, easily interpretable confidence intervals for the required effect sizes. In regulatory toxicology not only the difference of means is relevant as effect size \cite{Szoecs2015}, such as proportions or counts. Therefore, the Williams test is available for difference \cite{Hothorn2010}, risk ratio or odds ratio of proportions \cite{Hothorn2016}, ratio-to-control estimates \cite{Hothorn2011}, the nonparametric relative effect sizes \cite{Konietschke2012}, hazard rates \cite{Herberich2012}, multiple endpoints \cite{Hasler2012}, heteroscedastic error terms \cite{Herberich2010} and poly-k-adjusted tumor rates \cite{Schaarschmidt2008}).\\

\subsection{Factor laboratory: The total mean multiple contrast test}
Qualitative levels of a factor (without order restriction) are commonly compared by Tukey's all-pairs comparison procedure \cite{Tukey1953}.  Their numerous, exact $(l*(l-1)/2)*2$ comparisons make this procedure conservative and especially difficult to interpret. One or more deviating laboratories can also be identified by total mean comparisons \cite{Pallmann2016}, but with only $l$ to be interpreted. For $l=4$ laboratories the contrast matrix is (in a balanced design) simply
 \begin{table}[htbp]
            \begin{tabular}{ r c c c c }
         $c_i$ & $Lab_A$ & $Lab_B$ & $Lab_C$ &$Lab_D$ \\
        $c_a$ & -1 & 1/3   & 1/3 & 1/3   \\
        $c_b$ & 1/3 & -1 & 1/3 & 1/3  \\
				$c_c$ & 1/3 & 1/3 & -1 &1/3  \\
				$c_d$ & 1/3 & 1/3 & 1/3 &-1  \\
				\end{tabular}
        \end{table}

\subsection{Interaction dose-by-laboratory: Williams-by-total mean interaction contrast test}
When reformulating the common ANOVA interaction model: $y_{klj}=\mu+\alpha_k+\beta_l+ (\alpha\beta)_{kl}+\epsilon_{klj}$  (with $j$ replicates) into a cell means model $y_{klj}=\mu_{kl}+\epsilon_{klj}$ the null hypothesis of interaction can be written:\\
 $H_0: (\mu_{kl}- \mu_{kl'})-(\mu_{k'l}- \mu_{k'l'})=0$ for all 	$(k,k')$ and $(l,l')$ \cite{Kitsche2015} as a special form of product-type interactions \cite{Gabriel1973}. Using the both contrast matrices for Williams-type between doses $C_{Dose}$ and between laboratories total mean $C_lab$, the interaction contrast is the Kronecker product $C_{Dose,lab}=C_{Dose} \otimes C_{lab}$. This can easily be realized for different two-way contrast types with the CRAN-package \textit{statint} \cite{statint2014}.  Explicit for the above Ames assay example Table \ref{tab:IAS} contains the 36 individual contrasts for Williams-by-total mean interaction with k=7 $([0, 0.015625, 0.03125, 0.0625,0.125,0.25,0.5]$ and l=6 $[Lab_1, Lab_3, Lab_4, Lab_5, Lab_6, Lab_7])$: 

\begin{table}[ht]
\centering
\begingroup\tiny
\begin{tabular}{rll}
  \hline
 No. & IA-contrast lab:dose abbreviation & Contrast type \\ 
  \hline
1 & ((1 - 2,3,4,5,6,7):0.5) - ((1 - 2,3,4,5,6,7):0) & (Lab1 vs. all Labs) by Williams contrast 1 $(0.5-C-)$ \\ 
  2 & ((1 - 2,3,4,5,6,7):0.25,0.5) - ((1 - 2,3,4,5,6,7):0) &(Lab1 vs. all Labs) by Williams contrast 2 $((0.5+0.25)/2-C-)$ \\ 
  3 & ((1 - 2,3,4,5,6,7):0.125,0.25,0.5) - ((1 - 2,3,4,5,6,7):0) & (Lab1 vs. all Labs) by Williams contrast 3 $((0.5+0.25+0.125)/3-C-)$ \\ 
  4 & ((1 - 2,3,4,5,6,7):0.0625,0.125,0.25,0.5) - ((1 - 2,3,4,5,6,7):0) & (Lab1 vs. all Labs) by Williams contrast 4 $((0.5+0.25+0.125+0.0625)/4-C-)$ \\ 
  5 & ((1 - 2,3,4,5,6,7):0.03125,0.0625,0.125,0.25,0.5) - ((1 - 2,3,4,5,6,7):0) & (Lab1 vs. all Labs) by Williams contrasts 5 $((0.5+0.25+0.125+0.0625+0.031)/5-C-)$ \\ 
  6 & ((1 - 2,3,4,5,6,7):0.015625,0.03125,0.0625,0.125,0.25,0.5) - ((1 - 2,3,4,5,6,7):0) & (Lab1 vs. all Labs) by Williams contrast 6 $((0.5+0.25+0.125+0.0625+0.031+0.015)/6-C-)$ \\ \hline

		7 & ((2 - 1,3,4,5,6,7):0.5) - ((2 - 1,3,4,5,6,7):0) & (Lab2 vs. all Labs) by Williams contrast 1 \\ 
  8 & ((2 - 1,3,4,5,6,7):0.25,0.5) - ((2 - 1,3,4,5,6,7):0) & (Lab2 vs. all Labs) by Williams contrast 2 \\ 
  9 & ((2 - 1,3,4,5,6,7):0.125,0.25,0.5) - ((2 - 1,3,4,5,6,7):0) & (Lab2 vs. all Labs) by Williams contrast 3 \\ 
  10 & ((2 - 1,3,4,5,6,7):0.0625,0.125,0.25,0.5) - ((2 - 1,3,4,5,6,7):0) & (Lab2 vs. all Labs) by Williams contrast 4 \\ 
  11 & ((2 - 1,3,4,5,6,7):0.03125,0.0625,0.125,0.25,0.5) - ((2 - 1,3,4,5,6,7):0) & (Lab2 vs. all Labs) by Williams contrast 5 \\ 
  12 & ((2 - 1,3,4,5,6,7):0.015625,0.03125,0.0625,0.125,0.25,0.5) - ((2 - 1,3,4,5,6,7):0) & (Lab2 vs. all Labs) by Williams contrast 6 \\ \hline

	13 & ((3 - 1,2,4,5,6,7):0.5) - ((3 - 1,2,4,5,6,7):0) & (Lab3 vs. all Labs) by Williams contrast 1 \\ 
14 & ((3 - 1,2,4,5,6,7):0.25,0.5) - ((3 - 1,2,4,5,6,7):0) & (Lab3 vs. all Labs) by Williams contrast 2 \\ 
  15 & ((3 - 1,2,4,5,6,7):0.125,0.25,0.5) - ((3 - 1,2,4,5,6,7):0) & (Lab3 vs. all Labs) by Williams contrast 3 \\ 
  16 & ((3 - 1,2,4,5,6,7):0.0625,0.125,0.25,0.5) - ((3 - 1,2,4,5,6,7):0) & (Lab3 vs. all Labs) by Williams contrast 4 \\ 
  17 & ((3 - 1,2,4,5,6,7):0.03125,0.0625,0.125,0.25,0.5) - ((3 - 1,2,4,5,6,7):0) & (Lab3 vs. all Labs) by Williams contrast 5 \\ 
  18 & ((3 - 1,2,4,5,6,7):0.015625,0.03125,0.0625,0.125,0.25,0.5) - ((3 - 1,2,4,5,6,7):0) & (Lab3 vs. all Labs) by Williams contrast 6 \\ \hline 
  
	19 & ((4 - 1,2,3,5,6,7):0.5) - ((4 - 1,2,3,5,6,7):0) & (Lab4 vs. all Labs) by Williams contrast 1 \\ 
  20 & ((4 - 1,2,3,5,6,7):0.25,0.5) - ((4 - 1,2,3,5,6,7):0) &(Lab4 vs. all Labs) by Williams contrast 2 \\ 
  21 & ((4 - 1,2,3,5,6,7):0.125,0.25,0.5) - ((4 - 1,2,3,5,6,7):0) & (Lab4 vs. all Labs) by Williams contrast 3\\ 
  22 & ((4 - 1,2,3,5,6,7):0.0625,0.125,0.25,0.5) - ((4 - 1,2,3,5,6,7):0) & (Lab4 vs. all Labs) by Williams contrast 4 \\ 
  23 & ((4 - 1,2,3,5,6,7):0.03125,0.0625,0.125,0.25,0.5) - ((4 - 1,2,3,5,6,7):0) & (Lab4 vs. all Labs) by Williams contrast 5 \\ 
  24 & ((4 - 1,2,3,5,6,7):0.015625,0.03125,0.0625,0.125,0.25,0.5) - ((4 - 1,2,3,5,6,7):0) & (Lab4 vs. all Labs) by Williams contrast 6 \\ \hline 
  
	25 & ((5 - 1,2,3,4,6,7):0.5) - ((5 - 1,2,3,4,6,7):0) & (Lab5 vs. all Labs) by Williams contrast 1 \\ 
  26 & ((5 - 1,2,3,4,6,7):0.25,0.5) - ((5 - 1,2,3,4,6,7):0) & (Lab5 vs. all Labs) by Williams contrast 2 \\ 
  27 & ((5 - 1,2,3,4,6,7):0.125,0.25,0.5) - ((5 - 1,2,3,4,6,7):0) & (Lab5 vs. all Labs) by Williams contrast 3 \\ 
  28 & ((5 - 1,2,3,4,6,7):0.0625,0.125,0.25,0.5) - ((5 - 1,2,3,4,6,7):0) & (Lab5 vs. all Labs) by Williams contrast 4 \\ 
  29 & ((5 - 1,2,3,4,6,7):0.03125,0.0625,0.125,0.25,0.5) - ((5 - 1,2,3,4,6,7):0) & (Lab5 vs. all Labs) by Williams contrast 5 \\ 
  30 & ((5 - 1,2,3,4,6,7):0.015625,0.03125,0.0625,0.125,0.25,0.5) - ((5 - 1,2,3,4,6,7):0) & (Lab5 vs. all Labs) by Williams contrast 6 \\ \hline 
  
	31 & ((6 - 1,2,3,4,5,7):0.5) - ((6 - 1,2,3,4,5,7):0) & (Lab6 vs. all Labs) by Williams contrast 1 \\ 
  32 & ((6 - 1,2,3,4,5,7):0.25,0.5) - ((6 - 1,2,3,4,5,7):0) & (Lab6 vs. all Labs) by Williams contrast 2	\\ 
  33 & ((6 - 1,2,3,4,5,7):0.125,0.25,0.5) - ((6 - 1,2,3,4,5,7):0) & (Lab6 vs. all Labs) by Williams contrast 3 \\ 
  34 & ((6 - 1,2,3,4,5,7):0.0625,0.125,0.25,0.5) - ((6 - 1,2,3,4,5,7):0) & (Lab6 vs. all Labs) by Williams contrast 4	\\ 
  35 & ((6 - 1,2,3,4,5,7):0.03125,0.0625,0.125,0.25,0.5) - ((6 - 1,2,3,4,5,7):0) & (Lab6 vs. all Labs) by Williams contrast 5 \\ 
  36 & ((6 - 1,2,3,4,5,7):0.015625,0.03125,0.0625,0.125,0.25,0.5) - ((6 - 1,2,3,4,5,7):0) & (Lab6 vs. all Labs) by Williams contrast 6 \\ \hline 
  
	37 & ((7 - 1,2,3,4,5,6):0.5) - ((7 - 1,2,3,4,5,6):0) & (Lab7 vs. all Labs) by Williams contrast 1 \\ 
  38 & ((7 - 1,2,3,4,5,6):0.25,0.5) - ((7 - 1,2,3,4,5,6):0) & (Lab7 vs. all Labs) by Williams contrast 2 \\ 
  39 & ((7 - 1,2,3,4,5,6):0.125,0.25,0.5) - ((7 - 1,2,3,4,5,6):0) & (Lab7 vs. all Labs) by Williams contrast 3 \\ 
  40 & ((7 - 1,2,3,4,5,6):0.0625,0.125,0.25,0.5) - ((7 - 1,2,3,4,5,6):0) & (Lab7 vs. all Labs) by Williams contrast 4 \\ 
  41 & ((7 - 1,2,3,4,5,6):0.03125,0.0625,0.125,0.25,0.5) - ((7 - 1,2,3,4,5,6):0) & (Lab7 vs. all Labs) by Williams contrast 5 \\ 
  42 & ((7 - 1,2,3,4,5,6):0.015625,0.03125,0.0625,0.125,0.25,0.5) - ((7 - 1,2,3,4,5,6):0) & (Lab7 vs. all Labs) by Williams contrast 6 \\ 
   \hline
\end{tabular}
\endgroup
\caption{Interaction contrast structure for Williams trend-by-total means}
	\label{tab:IAS} 
\end{table}
This table of this already ample example makes clear the advantages and disadvantages of this approach: on the one hand, 36 individual decisions instead of one single F-test, on the other hand, the individual decisions allow the statement between which laboratories and which trend contrasts there is similarity and where not. In terms of an union-intersection test (UIT), the approach is conservative, but not so extreme because of the high correlation between the contrasts. If you want to know which interactions are equivalent and which are not, use a combined intersection-union-by-union-intersection test (IUT-UIT): IUT for the equivalence tests in both directions, and UIT between the IA contrasts. It is also possible to achieve a global statement by using an IUT-IUT (then the IA-test at the marginal $\alpha=0.05$ level), where all individual tests should fulfill $p> 0.10$.

\subsection{A possible simplification: $(D_k-C-)$-by-total mean interaction contrasts}
Assuming a monotonic dose-response dependence, the entire William contrast can be replaced by $(D_k-C-)$ comparison in a simplified way. This is surprising, but $(D_k-C-)$ is the most important single contrast in the William test, is part of the corresponding closure test \cite{Hothorn2016} and part of the strict trend test according \cite{Lin2019}. This reduction especially simplifies the interpretation to only $k$ interaction contrasts. This simplification should be avoided if downturns at higher dose(s) are possible or if you are more interested in comparing the NOAEC instead of a trend itself. The simplification becomes clear looking on Table \ref{tab:IAShigh} for the above Ames assay example: only 7 contrasts, claiming for local equivalence for laboratories $[1,2,3,4,5,6]$ and but not for global equivalence because laboratory 7 behaves borderline non-equivalent.
\begin{table}[ht]
\centering
\begingroup\tiny
\begin{tabular}{rlr}
  \hline
 Number& Interaction contrast & p-value \\ 
  \hline
1 & ((1 - 2,3,4,5,6,7):0.5) - ((1 - 2,3,4,5,6,7):0) & 0.985 \\ 
  2 & ((2 - 1,3,4,5,6,7):0.5) - ((2 - 1,3,4,5,6,7):0) & 0.410 \\ 
  3 & ((3 - 1,2,4,5,6,7):0.5) - ((3 - 1,2,4,5,6,7):0) & 0.999 \\ 
  4 & ((4 - 1,2,3,5,6,7):0.5) - ((4 - 1,2,3,5,6,7):0) & 0.999 \\ 
  5 & ((5 - 1,2,3,4,6,7):0.5) - ((5 - 1,2,3,4,6,7):0) & 0.584 \\ 
  6 & ((6 - 1,2,3,4,5,7):0.5) - ((6 - 1,2,3,4,5,7):0) & 0.120 \\ 
  7 & ((7 - 1,2,3,4,5,6):0.5) - ((7 - 1,2,3,4,5,6):0) & 0.100 \\ 
   \hline
\end{tabular}
\endgroup
\caption{Compound 4: interaction contrast for C vs. $D_k$ by grand means between labs- S9+}
	\label{tab:IAShigh}  
\end{table}

\subsection{A modification}
Variance heterogeneity often occurs in these assays and their ignorance can lead to significant bias. Therefore you should use the sandwich variance estimator (see below using vcov=sandwich argument in the function \textit{glht()}) \cite{Herberich2010}.

\section{Evaluation of the example using the CRAN packages \textit{statint} and \textit{multcomp}}
The following R-code can be used for the Ames assay example for the data object \textit{A984p}:
\scriptsize
\begin{verbatim}
library(multcomp); library(statint); library(sandwich)
InteractionContrastsP984t <- iacontrast(fa=A984p$lab,fb=A984p$Conc,
                                       typea="totalMean", typeb="Williams")
A984p$labConct <- InteractionContrastsP984t$fab
CellMeansModelP984t <- lm(Trans ~ labConct-1, data=A984p) # cell means model
MultTestP984t <- glht(model=CellMeansModelP984t,
                 linfct = mcp(labConct=InteractionContrastsP984t$cmab), vcov=sandwich)
piap984t<-summary(MultTestP984t)						#calculating adjusted p-values
\end{verbatim}

\begin{table}[ht]
\centering
\begingroup\tiny
\begin{tabular}{rlr}
  \hline
 Number & Interaction contrast & p-value \\ 
  \hline
1 & ((1 - 2,3,4,5,6,7):0.5) - ((1 - 2,3,4,5,6,7):0) & 0.9994 \\ 
  2 & ((1 - 2,3,4,5,6,7):0.25,0.5) - ((1 - 2,3,4,5,6,7):0) & 0.999 \\ 
  3 & ((1 - 2,3,4,5,6,7):0.125,0.25,0.5) - ((1 - 2,3,4,5,6,7):0) & 0.999 \\ 
  4 & ((1 - 2,3,4,5,6,7):0.0625,0.125,0.25,0.5) - ((1 - 2,3,4,5,6,7):0) & 0.9698 \\ 
  5 & ((1 - 2,3,4,5,6,7):0.03125,0.0625,0.125,0.25,0.5) - ((1 - 2,3,4,5,6,7):0) & 0.8706 \\ 
  6 & ((1 - 2,3,4,5,6,7):0.015625,0.03125,0.0625,0.125,0.25,0.5) - ((1 - 2,3,4,5,6,7):0) & 0.9230 \\ 
  7 & ((2 - 1,3,4,5,6,7):0.5) - ((2 - 1,3,4,5,6,7):0) & 0.5901 \\ 
  8 & ((2 - 1,3,4,5,6,7):0.25,0.5) - ((2 - 1,3,4,5,6,7):0) & 0.4831 \\ 
  9 & ((2 - 1,3,4,5,6,7):0.125,0.25,0.5) - ((2 - 1,3,4,5,6,7):0) & 0.8529 \\ 
  10 & ((2 - 1,3,4,5,6,7):0.0625,0.125,0.25,0.5) - ((2 - 1,3,4,5,6,7):0) & 0.8546 \\ 
  11 & ((2 - 1,3,4,5,6,7):0.03125,0.0625,0.125,0.25,0.5) - ((2 - 1,3,4,5,6,7):0) & 0.9306 \\ 
  12 & ((2 - 1,3,4,5,6,7):0.015625,0.03125,0.0625,0.125,0.25,0.5) - ((2 - 1,3,4,5,6,7):0) & 0.9719 \\ 
  13 & ((3 - 1,2,4,5,6,7):0.5) - ((3 - 1,2,4,5,6,7):0) & 0.999 \\ 
  14 & ((3 - 1,2,4,5,6,7):0.25,0.5) - ((3 - 1,2,4,5,6,7):0) & 0.999 \\ 
  15 & ((3 - 1,2,4,5,6,7):0.125,0.25,0.5) - ((3 - 1,2,4,5,6,7):0) & 0.9990 \\ 
  16 & ((3 - 1,2,4,5,6,7):0.0625,0.125,0.25,0.5) - ((3 - 1,2,4,5,6,7):0) & 0.9887 \\ 
  17 & ((3 - 1,2,4,5,6,7):0.03125,0.0625,0.125,0.25,0.5) - ((3 - 1,2,4,5,6,7):0) & 0.999 \\ 
  18 & ((3 - 1,2,4,5,6,7):0.015625,0.03125,0.0625,0.125,0.25,0.5) - ((3 - 1,2,4,5,6,7):0) & 0.999 \\ 
  19 & ((4 - 1,2,3,5,6,7):0.5) - ((4 - 1,2,3,5,6,7):0) & 0.999 \\ 
  20 & ((4 - 1,2,3,5,6,7):0.25,0.5) - ((4 - 1,2,3,5,6,7):0) & 0.999 \\ 
  21 & ((4 - 1,2,3,5,6,7):0.125,0.25,0.5) - ((4 - 1,2,3,5,6,7):0) & 0.999 \\ 
  22 & ((4 - 1,2,3,5,6,7):0.0625,0.125,0.25,0.5) - ((4 - 1,2,3,5,6,7):0) & 0.999 \\ 
  23 & ((4 - 1,2,3,5,6,7):0.03125,0.0625,0.125,0.25,0.5) - ((4 - 1,2,3,5,6,7):0) & 0.999 \\ 
  24 & ((4 - 1,2,3,5,6,7):0.015625,0.03125,0.0625,0.125,0.25,0.5) - ((4 - 1,2,3,5,6,7):0) & 0.999 \\ 
  25 & ((5 - 1,2,3,4,6,7):0.5) - ((5 - 1,2,3,4,6,7):0) & 0.7772 \\ 
  26 & ((5 - 1,2,3,4,6,7):0.25,0.5) - ((5 - 1,2,3,4,6,7):0) & 0.8020 \\ 
  27 & ((5 - 1,2,3,4,6,7):0.125,0.25,0.5) - ((5 - 1,2,3,4,6,7):0) & 0.9052 \\ 
  28 & ((5 - 1,2,3,4,6,7):0.0625,0.125,0.25,0.5) - ((5 - 1,2,3,4,6,7):0) & 0.9928 \\ 
  29 & ((5 - 1,2,3,4,6,7):0.03125,0.0625,0.125,0.25,0.5) - ((5 - 1,2,3,4,6,7):0) & 0.9828 \\ 
  30 & ((5 - 1,2,3,4,6,7):0.015625,0.03125,0.0625,0.125,0.25,0.5) - ((5 - 1,2,3,4,6,7):0) & 0.9999 \\ 
  31 & ((6 - 1,2,3,4,5,7):0.5) - ((6 - 1,2,3,4,5,7):0) & 0.1843 \\ 
  32 & ((6 - 1,2,3,4,5,7):0.25,0.5) - ((6 - 1,2,3,4,5,7):0) & 0.2064 \\ 
  33 & ((6 - 1,2,3,4,5,7):0.125,0.25,0.5) - ((6 - 1,2,3,4,5,7):0) & 0.7596 \\ 
  34 & ((6 - 1,2,3,4,5,7):0.0625,0.125,0.25,0.5) - ((6 - 1,2,3,4,5,7):0) & 0.8450 \\ 
  35 & ((6 - 1,2,3,4,5,7):0.03125,0.0625,0.125,0.25,0.5) - ((6 - 1,2,3,4,5,7):0) & 0.9857 \\ 
  36 & ((6 - 1,2,3,4,5,7):0.015625,0.03125,0.0625,0.125,0.25,0.5) - ((6 - 1,2,3,4,5,7):0) & 0.9952 \\ 
  37 & ((7 - 1,2,3,4,5,6):0.5) - ((7 - 1,2,3,4,5,6):0) & 0.1511 \\ 
  38 & ((7 - 1,2,3,4,5,6):0.25,0.5) - ((7 - 1,2,3,4,5,6):0) & 0.4369 \\ 
  39 & ((7 - 1,2,3,4,5,6):0.125,0.25,0.5) - ((7 - 1,2,3,4,5,6):0) & 0.3633 \\ 
  40 & ((7 - 1,2,3,4,5,6):0.0625,0.125,0.25,0.5) - ((7 - 1,2,3,4,5,6):0) & 0.4856 \\ 
  41 & ((7 - 1,2,3,4,5,6):0.03125,0.0625,0.125,0.25,0.5) - ((7 - 1,2,3,4,5,6):0) & 0.7422 \\ 
  42 & ((7 - 1,2,3,4,5,6):0.015625,0.03125,0.0625,0.125,0.25,0.5) - ((7 - 1,2,3,4,5,6):0) & 0.8666 \\ 
   \hline
\end{tabular}
\endgroup
\caption{Compound 4: interaction contrast for Williams trend-by grand means between labs- S9+} 
	\label{tab:IAWil}  

\end{table}

\normalsize
The related adjusted p-values for this IUT-UIT test in Table \ref{tab:IAWil} are predominantly large, which indicate equivalence, even global equivalence. Only one contrast, No. 37 shows with $p=0.1511$ a tendency of non-equivalence; but the single comparison $(D_k-C-)$ should not be relevant for the Ames assay because the decision is based on the lower doses. This p-value of $0.1511$ is part of the comparisons of Lab 7 against all others, which are not as large as the others (but still equivalent). Notice, Lab 7 used a different source of bacteria revealing a higher spontaneous reversion rate.\\

\begin{figure}
	\centering
		\includegraphics[width=0.692\textwidth]{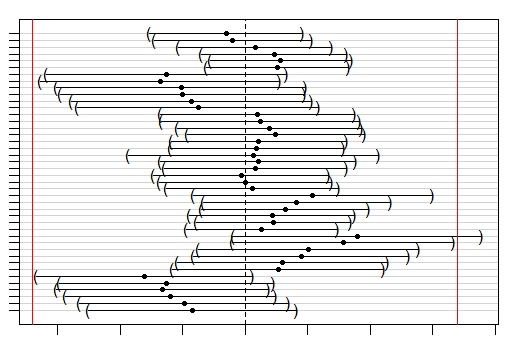}
		\caption{42 interaction contrasts: compatible simultaneous confidence intervals}
	\label{fig:CII}
\end{figure}

 Even more informative is the plot of the simultaneous two-sided $(1-2\alpha)$ confidence intervals in Figure \ref{fig:CII}, demonstrating the problems with Lab 7 in the scale of transformed revertant count differences. Notice, using the IUT-IUT approach with marginal levels, global equivalence can be claimed for this selected example.

Notice, the (two-sided) p-value for the common F-test on dose-by-laboratory interaction is $0.10$ indicating that an interaction cannot be excluded. However, it is not clear for which laboratories and which dose comparisons - a nice proof of the usefulness of the above IA-contrast method in this example.

\section{Summary}
Similarity of multiple dose-response curves in interlaboratory studies in regulatory toxicology can be demonstrated using dose-by-laboratory interaction contrasts \cite{Hothorn2003, Hothorn2016}. To illustrate a trend of the dose-response curve, Williams-by-Laboratory interaction contrasts are proposed here. With help of the CRAN packages \textit{statint} and \textit{multcomp} the estimation of adjusted p-values or compatible simultaneous confidence intervals is relatively easy. The interpretation in terms of global or partial equivalence (similarity) is challenging, but impressive.

\footnotesize
\bibliographystyle{plain}

\normalsize

\end{document}